# Evidences of a threshold system as the source for magnetic storms detected on Earth's surface

Andrés R. R. Papa[1,2], Luiz M. Barreto[1] & Ney A. B. Seixas[1]

[1]*Observatório Nacional, MCT/ON, Rua General José Cristino 77, São Cristovão, Rio de Janeiro, 20921-400RJ, BRAZIL*

[2]*Department of Mathematics and Computation, State University of Rio de Janeiro, Estrada Resende –Riachuelo s/n, Morada da Colina, Resende, 27523-000 RJ, BRAZIL*

*'These authors contributed equally to this work'*

**Threshold systems appear to underlie the global behaviour of physical phenomena very unlike at a first look. The usual experimental fingerprint of threshold system grounded phenomena is the presence of power laws. Experimental evidence has been found, for example, in superconductor's vortex avalanches[1], sand piles[2], the brain[3], $^4$He superfluitdity[4] and earthquakes[5]. Double power-laws have been found in social networks[6], in the luminosity of some galactic nuclei[7] and, very recently, in solar flares[8], among others. Here we show evidences that point to a threshold system as the source for magnetic storms detected on the Earth. We based our analysis on series of data acquired during many years in the network of magnetic observatories of the National Observatory (Brazil). In particular we focused our attention on October 2000 month of the Vassouras Observatory (RJ, Brazil), which have been active since 1915. We have found both power laws and double power laws for some relevant distributions.**

It is well known that any of the components of the magnetic field measured on the Earth's surface presents characteristic frequencies (corresponding to time periods of 24 hours and harmonics) owing to the rotation of Earth. There are also present other factors (for example, ionospheric influences). However, in this approach we were concerned just with the elimination of well-known frequencies (tidal corrections and electrojets corrections, for instance, were not considered).

In Fig. 1a we show the temporal dependence of the *H* component of the magnetic field measured on the ground during a day (October 14, 2000). We have used the *H* component because, due to the Vassouras Observatory low geomagnetic latitude, it is almost equal to the total field *F*. The rate of registered data was one point per minute giving 44640 lectures for the whole month.

To extract a signal independent of daily influences we closely followed a method developed in previous works[9,10]. There was proposed a computational procedure for the automatic determination of the *K*-index for magnetic activity, instead of the classical hand-made one. It is based on the application of a Fast Fourier Transform (FFT) using a Butterworth's low-pass filter with a 12 *dB/octave* gradient and a variable cut-off frequency given by the expression $f_C = \delta_{obs}/A_{pp}$, where $\delta_{obs}$ is a constant (for the Vassouras Observatory it is 0.26) and $A_{pp}$ is the peak-to-peak amplitude in a daily register. The main difference for the present study is that the total time for analysis must be as long as possible in order to have several decades (if allowed by the nature of the phenomenon) for the statistical description of stormy and calm periods.

It is also shown in Fig. 1a the FFT after filtering and returning to time domain processes. All the irregular behaviour, that is actually the subject we are interested on, has been erased off.

Next it was calculated the difference *ΔH* between the filtered and unfiltered signals. The result for one day is shown in Fig. 1b. Through this procedure we isolated the spike-like characteristic. To study the frequency distribution of amplitudes, of stormy periods and of first return times it was previously obtained the absolute value of the difference between data of the types presented in Fig. 1a ($|\Delta H|$). The objective was to avoid artificial bisections of single stormy periods. It is know that during those periods the magnetic field components change the sign continuously.

Figure 2 shows the frequency distribution of stormy periods. It was calculated over a period of one month with magnetic field measures every 1 minute. For Fig. 2 it was assumed a cut-off of 0.9 *nT* (*i.e.*, it was considered as "a storm" every continuous activity above 0.9 *nT*). We have also represented in Fig. 2 a straight line of slope equal to –2.61 that helps to call the attention to the power-law-like distribution for stormy periods. For low duration time values ($t < 20$ *minutes*) another incipient power-law seems to be present. However, we prefer not to claim that because it extends for less than a decade (and is probably partially affected by the 1 point/minute register rate).

Motivated for these facts and looking for more deep details we have calculated the frequency distribution of the *ΔH* component of the magnetic field. The results are shown in Fig. 3. Now, there are two, better defined, apparent power law regimes. For values up to 20 *nT* the slope is –1.86, for higher values it was obtained a value of –3.70 for the slope. We are bent to think that the double power-law originates from two very different networks of concomitant phenomena at the Sun surface. Future modelling works will help to explain this result. Double power-laws have been found very recently[8] for the X-rays energy distribution in solar flares. Both findings probably have a subtle connection through an energy relation. In any case special attention must be paid to the strange coincidence of one of the slopes be very close to a half of the other.

Finally, it is presented in Fig. 4 the frequency distribution for first return times (duration of calm periods), measured during the same monthly period, with magnetic field measurements every minute. For Fig. 4 it was also assumed a cut-off of 0.9 *nT* 4(*i.e.*, it was considered as "a calm period" every continuous activity below 0.9 *nT*). It is also depicted in Fig. 4 a straight line of slope equal to –2.17 that helps to call the attention to the power-law-like distribution of first return times.

The results shown in figures 2, 3 and 4 are a strong evidence of power laws for the frequency distribution of peak highs and of stormy and calm periods in magnetic measurements on the Earth surface. This might give support to more detailed treatments of available data in future works and to the search and construction of models to appropriately represent this fact. The first published evidence in this direction that we have found was published more than forty years ago[11] and presented as the harmonic analysis of the around-the-world magnetic profile. But at that time it was not clear all the consequences that 1/f noise[5] could carry with it.

Resuming, we have found for the frequency distributions of stormy periods duration, of calm periods duration and of peak highs, laws of the type:

$$f(q) = cq^d \qquad (1)$$

where *f(q)* is the frequency distribution on the variable *q*, *c* is some proportionality constant and *d* the exponent of the power law. When *q* represents the duration of stormy periods $d = -2.61$ and when it represents first return times (duration of calm periods) $d = -2.17$. For the distribution of peak highs we have found two laws of the type shown in Equation (1). Up to, approximately, 20 *nT*, $d = -1.86$. For higher values $d = -3.70$.

Both frequency distributions, of stormy periods and of first return times, point to a threshold system as the source of magnetic storms measured on the Earth. Models for the Sun (which is the ground for this system) should describe this particularity. They should also describe the double power-law present in the amplitude distribution (and probably in the duration of stormy periods). Independently of the success of those models in explaining the results here presented, the present work shows that together with earthquakes and many others systems known to be threshold driven ones there should be considered the magnetic storms.

**Competing Interests statement** The authors declare that they have not competing financial interests.

**Correspondence** and requests for materials should be addressed to A.R.R.P. (e-mail: papa@on.br).

Figure 1 – Typical daily time dependence of the H component. **a**. Time dependence of the magnetic field H measured on the Earth's surface during one day. Actually there are several day classifications depending on the magnetic activity: calm days, moderate days and stormy days. The day here presented (October 14, 2000) was a moderate one. The black irregular curve corresponds to natural data, the red soft curve to data after FFT processing. **b**. Difference between the signals in Fig. 1a (*ΔH*). Note the spike-like shape. Note also the different coordinate scale in relation to Fig. 1a. On this type of data set were performed the final data analysis.

Figure 2 – Storms duration distribution. The red straight line is a guide for the eye. It has slope –2.61. We have not risked a second power law possibility (for lower times) because it stands for less than a decade and, fundamentally, because there are few points in that range.

Figure 3 – Distribution of $\Delta H$. The straight lines are guides for the eye. A double regime seems to be present. Although each of the power laws extends for only a decade (or less) the number of experimental points brings support to this affirmation. The red straight line has a slope of –1.86 while the value for the blue one is –3.70.

Figure 4 – Frequency distribution of first return times (calm periods). The red straight line is a guide for the eye. It has slope –2.17. Even though there are few points in the 10-100 *minutes* decade (where the linear tendency is clear), the distribution of points for times greater than 100 *minutes* supports the hypothesis of the existence of a power law.

**FIGURE 1**

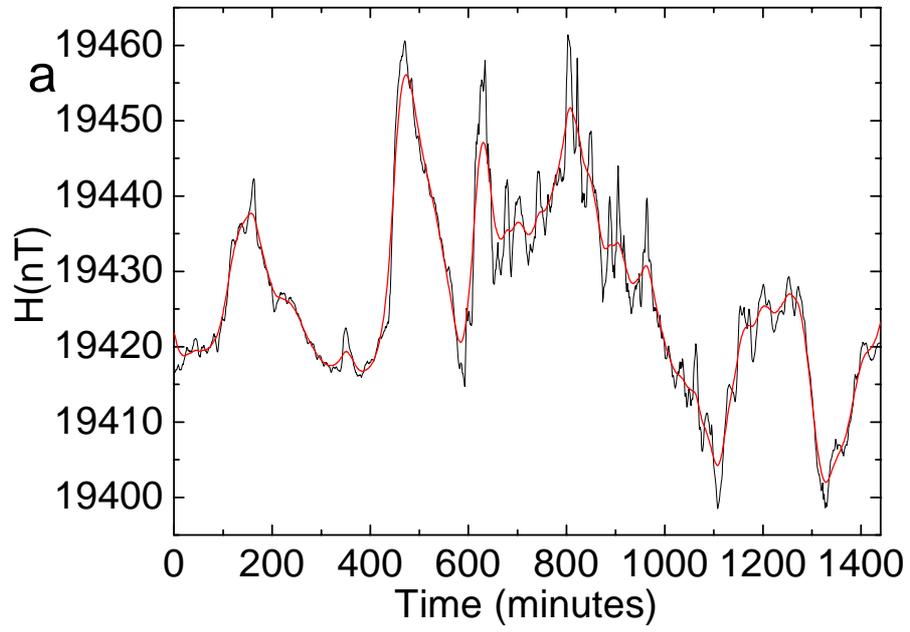

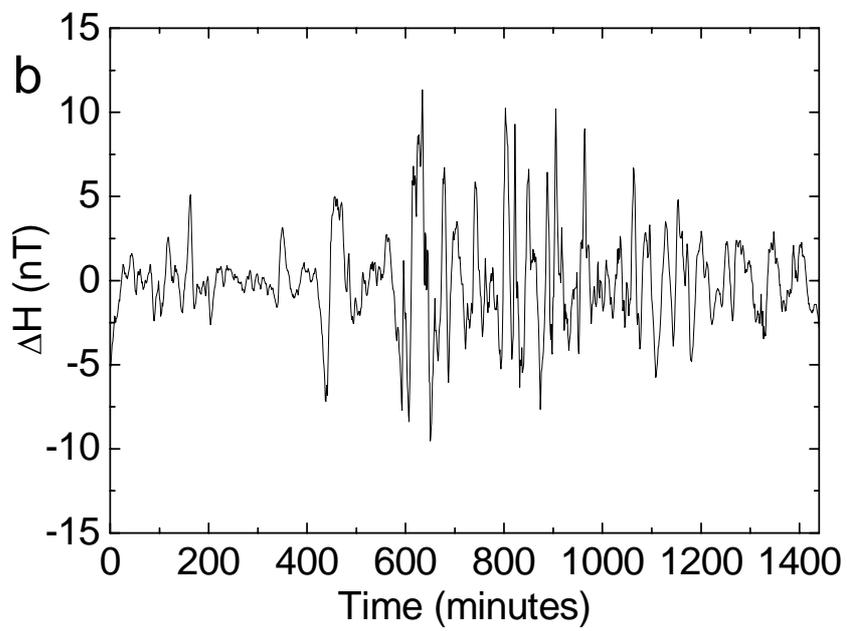

**FIGURE 2**

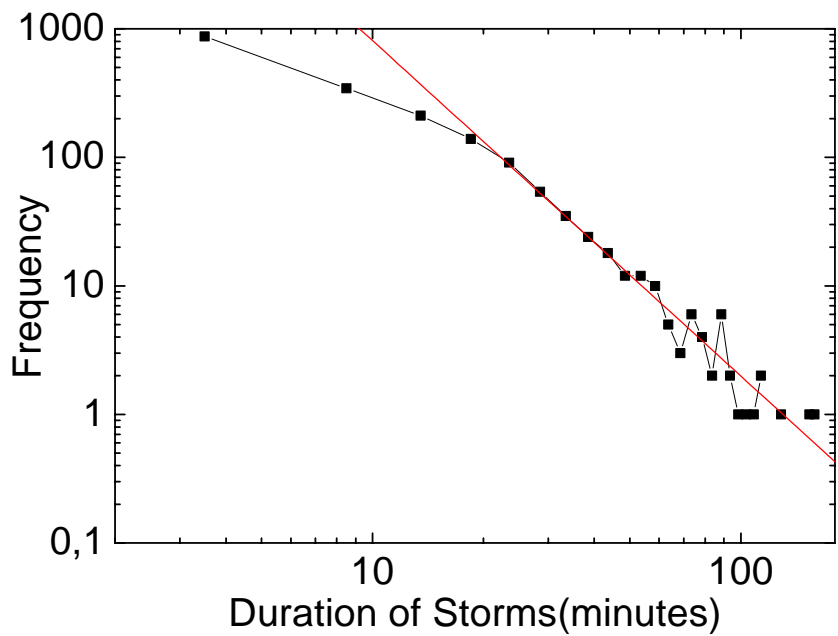

**FIGURE 3**

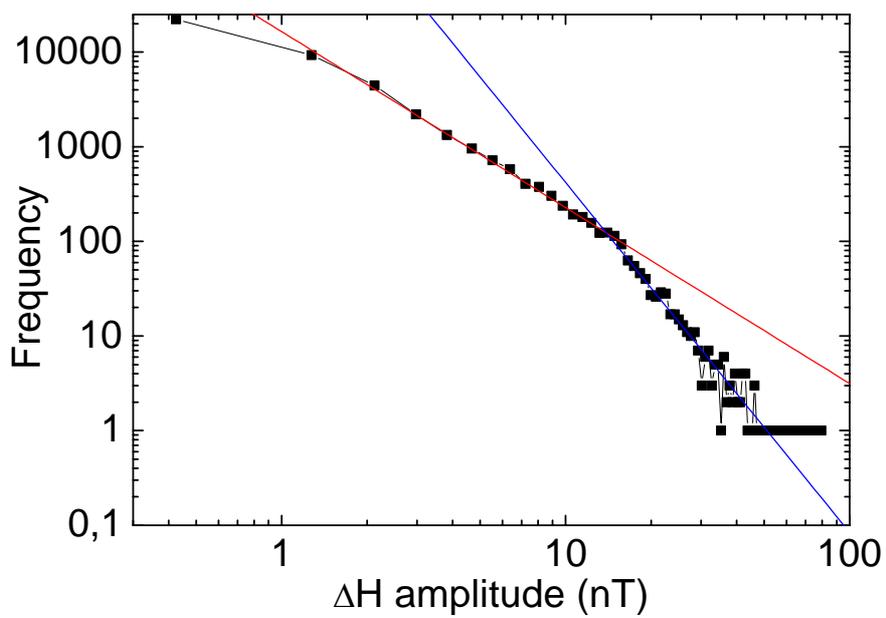

**FIGURE 4**

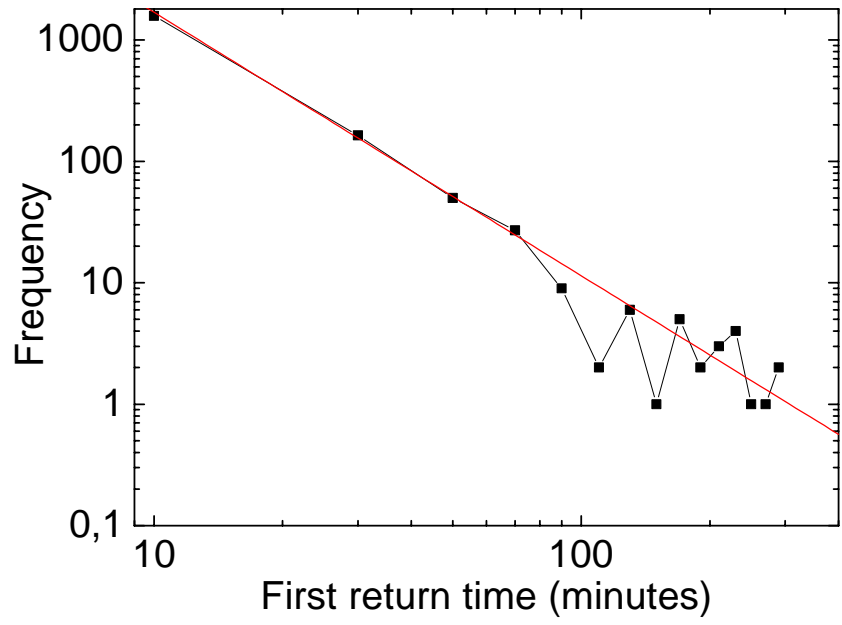